\documentclass[prd,showpacs,showkeys,floatfix,nofootinbib,  
                  preprint,12pt,tightenlines,fleqn]{revtex4} 

\usepackage{amsmath,amssymb,revsymb,graphicx,dcolumn}

\newcommand{\beq}{\begin{equation}}
\newcommand{\eeq}{\end{equation}}
\newcommand{\beqa}{\begin{eqnarray}}
\newcommand{\eeqa}{\end{eqnarray}}

\newcommand{\half}{{\textstyle \frac{1}{2}}}    

\begin{document}

\noindent Physical Review D 77, 085015 (2008) \hfill
          arXiv:0711.3170v6 [gr-qc]\newline\vspace*{1\baselineskip}
\title[Self-tuning vacuum variable ...]
        {Self-tuning vacuum variable and cosmological constant
        \vspace*{.5\baselineskip}}
\author{F.R.~Klinkhamer}
\email{frans.klinkhamer@physik.uni-karlsruhe.de}
\affiliation{Institute for Theoretical Physics,\\
University of Karlsruhe (TH),\\
76128 Karlsruhe, Germany}
\author{G.E. Volovik}
\email{volovik@boojum.hut.fi}
\affiliation{Low Temperature Laboratory,
Helsinki University of Technology\\
P.O. Box 2200, FIN-02015 HUT, Finland\\
and\\
L.D. Landau Institute for Theoretical Physics,
Russian Academy of Sciences, Kosygina 2, 119334 Moscow, Russia\\}

\begin{abstract}
\vspace*{.25\baselineskip}\noindent
A spacetime-independent variable is introduced which characterizes
a Lorentz-invariant self-sustained quantum vacuum.
For a perfect (Lorentz-invariant) quantum vacuum,
the self-tuning of this variable
nullifies the effective energy density
which enters the low-energy gravitational field equations.
The observed small but nonzero value of the cosmological constant
may then be explained as corresponding to the
effective energy density of an imperfect quantum vacuum
(perturbed by, e.g., the presence of thermal matter).
\end{abstract}

\pacs{11.30.Cp, 04.20.Cv, 98.80.Jk, 95.36.+x}
\keywords{Lorentz invariance, general relativity, cosmology, dark energy}
\maketitle

\section{Introduction}
\label{sec:introduction}

A first step towards the elusive theory of ``quantum gravity''
\cite{Rovelli2004,Kiefer2007} is to identify the microscopic constituents
of space \cite{Hu2005,Volovik2006,Klinkhamer2007,Oriti2007}.
A next step which may be strongly correlated with the first step
is to properly describe the dynamics of these constituents.

In the last few years, a new approach to the above mentioned
program has emerged, which has been
driven by both experimental and theoretical considerations.
Astronomical observations suggest the existence of a
cosmological constant \cite{Einstein1917} or vacuum energy density
\cite{Bronstein1933,Zeldovich1967,Weinberg1972,Veltman1976}
with a typical energy scale of the order of  $10^{-3}\;\text{eV}$
\cite{deBernardis2000,Hinshaw2006,Riess2006}.
But current theories do not provide a
good symmetry explanation for the smallness of this value
\cite{Weinberg1988,SahniStarobinsky2000,Padmanabhan2002,Nobbenhuis2004,Polchinski2006}.
This last observation suggests that
power-divergent quantum corrections to the vacuum energy density
would not be regulated by a dynamical principle \cite{Giudice2007}.
Instead, the vacuum energy density would have to be regulated
by a more general principle.

The aim of this article is to present \emph{thermodynamic} arguments for
the behavior of the effective vacuum energy density
which enters the low-energy gravitational field equations.
These arguments will be formulated in the context of emergent-gravity
models \cite{Bjorken2001,Laughlin2003,FroggattNielsen2005,Volovik2007}
but may have a more general applicability.

Let us give, right from the start, a heuristic
argument \cite{Dreyer2006} in favor of the emergent-symmetry approach
to the cosmological constant problem.
Namely, if gravitation would be a truly fundamental interaction,
it would be hard to understand why the energies stored
in the quantum vacuum would not gravitate at all
\cite{Weinberg1988,SahniStarobinsky2000,Padmanabhan2002,Nobbenhuis2004,Polchinski2006}.
If, however, gravitation would be only a low-energy effective interaction
\cite{Bjorken2001,Laughlin2003,FroggattNielsen2005,Volovik2007},
it could be that the corresponding gravitons as quasiparticles
do not feel \emph{all} microscopic degrees of freedom
(gravitons would be analogous to small-amplitude waves
at the surface of the ocean)
and that the gravitating effect of the total vacuum energy density would be
effectively \emph{tuned away.} In fact, it is at this point that elementary
thermodynamic arguments \cite{Einstein1907} might be useful,
even if the details of the underlying microscopic physics
(the ocean in the above analogy) remain unknown.

A brief outline of this article runs as follows.
In Sec.~\ref{sec:Thermodynamics-vacuum-variable},
a particular Lorentz-invariant vacuum variable is introduced
which allows us to discuss thermodynamic properties of the vacuum
such as stability, compressibility, and
thermodynamic response to perturbations.
In Sec.~\ref{sec:NullificationLambda},
it is argued that this vacuum variable adjusts itself so as to nullify
the relevant vacuum energy density which contributes to the effective
gravitational fields equations and the effective cosmological
constant of a perfect quantum vacuum would be zero.
In Sec.~\ref{sec:Nonzero-vacuum-energy-density},
it is shown how the presence of thermal matter shifts the value of this effective
vacuum energy density away from zero (the same happens with other types of
perturbations, such as the existence of a spacetime boundary of the volume
considered).
In Sec.~\ref{sec:Nature-and-origin},
the possible origin of the vacuum variable is discussed in general terms,
but two specific examples are also presented
(one with a four-form field strength and another with a four-velocity field).
In Sec.~\ref{sec:Summary},
our results are summarized and an outlook is given.

\section{Vacuum variable and thermodynamics}
\label{sec:Thermodynamics-vacuum-variable}
\subsection{New variable for the Lorentz-invariant quantum vacuum}
\label{sec:Variable}

The quantum vacuum (or ``aether'' in the old terminology
\cite{Maxwell1892,Lorentz1927,Einstein1920})
is a medium of a peculiar nature. The most important property of the
quantum vacuum is its Lorentz invariance.
Current experimental bounds on Lorentz-violating modifications of
Maxwell theory at the level of
$10^{-18}$ or better \cite{KosteleckyMewes2002,KlinkhamerRisse2007}
suggest that a hypothetical Lorentz-violating energy scale $E_\text{LV}$
\cite{KlinkhamerVolovik2005JETPL} exceeds
the Planck energy scale $E_\text{Planck}\equiv \sqrt{\hbar\, c^5/G}$
by at least 9 orders of magnitude, for Lorentz-violating
modifications of order $(E_\text{Planck}/E_\text{LV})^2$.
The Lorentz invariance of the quantum vacuum  imposes constraints on its
kinematics and dynamics. According to Einstein \cite{Einstein1920},
``this aether may not be thought of as
endowed with the quality characteristic of ponderable media''
and ``the idea of motion may not be applied to it.''
These considerations suggest that the momentum density of
the quantum vacuum is strictly zero. With Lorentz invariance
playing a crucial role, it will be natural to use units with $c=1$
in the rest of this article.
The same holds for quantum mechanics and we also set $\hbar=1$.

Another important property of the quantum vacuum is
the apparent discrepancy between the magnitude of the theoretically expected
contributions to the vacuum energy density and the experimentally observed value
of the effective vacuum energy density. The formal one-loop contribution
is proportional to $M^4$, where $M$ is the characteristic mass scale of
the particles and fields which make up the vacuum and where a
relativistically-invariant regularization must be used in order to recover
a Lorentz-invariant equation of state for the quantum vacuum (see, e.g.,
Appendices II and VIII of the well-known review \cite[(b)]{Zeldovich1967}
and the more recent article \cite{Akhmedov2002}). The particular value
of $M$ depends on the detailed theory. If the vacuum is made solely
from the elementary particles and fields
of the electroweak standard model, the characteristic vacuum energy density
$\epsilon_\text{vac}$ is of order $M_W^4 \approx \big(10^{11}\,\text{eV}\big)^4$.
In the Frolov--Fursaev scheme \cite{FrolovFursaev1998} of Sakharov-like
induced gravity \cite{Sakharov1967}, the vacuum is
made from constituent bosonic and fermionic fields whose masses are of the
order of the Planck scale. This gives rise to a ``natural''  value
of the vacuum energy density $\epsilon_\text{vac}$ of order
$E^4_\text{Planck} \approx \big(10^{28}\,\text{eV}\big)^4$,
unless there is a special cancelation between the different species
caused by, for example, supersymmetry. For the case of supersymmetric
cancelations, the remaining vacuum energy is determined by the mass scale
at which supersymmetry is broken,
$M_\text{SUSYbreaking} \gtrsim  1\,\text{TeV} = 10^{13}\,\text{eV}$.
Anyway, regardless of which extension of the electroweak
standard model turns out to be correct,
the theoretical estimate of the vacuum energy,
$\epsilon_\text{vac}^\text{theo} \gtrsim \big(10^{11}\,\text{eV}\big)^4$,
is very much larger than the vacuum energy density indicated by
astronomical observations \cite{deBernardis2000,Hinshaw2006,Riess2006},
$\epsilon_\text{vac}^\text{obs} \approx \big(10^{-3}\,\text{eV}\big)^4$.

Still, this need not present an insurmountable problem. Having a zero value
(or almost zero value) of the relevant vacuum energy density
can be the property of a \emph{self-sustained medium},
that is, a medium with a definite macroscopic volume
even in the absence of an environment.
An example of such a self-sustained medium is a droplet of water
falling in empty space.

The observed near-zero value of the cosmological constant
compared to a Planck-scale value suggests that the quantum vacuum of our universe
belongs to this class of systems, namely, the class of self-sustained
media.\footnote{\label{ftn:selfsustained}A related suggestion is to
consider ``empty spacetime'' as a type of condensate \cite{Hu2005}.
Here, however, we require that such a condensate should be self-sustained,
i.e., that it should support an equilibrium state even in the absence
of an external environment. Conventional Bose--Einstein
condensates do not satisfy this requirement, since they exist
as equilibrium states only under applied external pressure.
It is to be emphasized that the analysis of the present article
does not rely on a condensed-matter-physics analogy
but aims to be self-contained, being guided only by
thermodynamics and Lorentz invariance.}
As any other medium of this kind, the equilibrium vacuum
would be homogeneous and extensive. The homogeneity assumption is
indeed supported by the observed flatness and
smoothness of our universe \cite{deBernardis2000,Hinshaw2006,Riess2006}.
The implication is that the energy of the equilibrium quantum
vacuum would be proportional to the volume considered.

Usually, a self-sustained medium is characterized by an
\emph{conserved extensive quantity}
whose total value determines the actual volume of the
system \cite{Pippard1964,LandauLifshitzStatMech1,Perrot1998}.
For this reason, we suggest that the quantum vacuum is \emph{also}
characterized by such a variable \cite{BjorkenPrivateComm}.
The Lorentz invariance of the quantum vacuum imposes,
however, strong constraints on the possible form this variable can take.
In particular, any global charge such as the fermionic charge $B-L$
of the standard model must be zero in the Lorentz-invariant vacuum.
(In the standard model, $B$ and $L$ are the baryon and lepton number,
with $B-L$ conserved and $B+L$ anomalously violated \cite{'tHooft1976}.)
The reason for allowing only $B-L=0$ in the Lorentz-invariant vacuum
is that $B-L$ is the integrated fourth component of the corresponding
4--current, so that $B-L$ transforms nontrivially under
Lorentz transformations if $B-L\neq 0$.

In order to be specific, we choose the vacuum variable to be a
symmetric tensor $q^{\mu\nu}(x)$ satisfying the following conservation law:
\begin{equation}
\nabla_\mu \,q^{\mu\nu}(x) =0~,
\label{eq:conservation}
\end{equation}
where $\nabla_\mu$ is the standard covariant derivative defined in terms
of the metric $g_{\mu\nu}(x)$, its inverse $g^{\mu\nu}(x)$, and their
derivatives \cite{Weinberg1972,Veltman1976}.
In a homogeneous vacuum, one has $q^{\mu\nu}(x)=q\,g^{\mu\nu}(x)$
with $q$ constant over space and time.
This new degree of freedom $q$ would be responsible for the equilibration
of the quantum vacuum: the equilibrium value of $q$ readjusts
if the vacuum is perturbed towards a new equilibrium state
(see Sec.~\ref{sec:Readjustment}). The quantum vacuum can now be considered
as a reservoir of trans-Planckian energies stored in the $q$--field
(see Sec.~\ref{sec:Nature-and-origin}
for further discussion on the possible origin of $q$).

As an elementary \emph{Gedankenexperiment}, take a large portion of
quantum vacuum which is supposed to be isolated from its
environment. This implies having vanishing external pressure, $P=0$.
For simplicity, consider only two  fields: the low-energy effective
matter field $\Psi(x)$ and the vacuum field $q(x)$.
Specifically, choose the classical field $\Psi(x)$ to be a complex
scalar field. The volume of the
isolated portion of quantum vacuum is variable, but its total ``charge''
$Q(t)\equiv \int d^3r~q(\mathbf{r},t)$ must be conserved,
$\mathrm{d}Q/\mathrm{d}t=0$. Throughout this article, we neglect
the possible spacetime dependence of $q$ and consider $q$ to be a genuine
thermodynamic variable.

The energy of this portion of quantum vacuum at fixed ``charge'' $Q=q\, V$
is then given by
\begin{equation}
E =\int d^3r~\epsilon\left(\Psi,q\right) =
\int d^3r~\epsilon\left(\Psi,Q/V\right)~.
\label{eq:P=0}
\end{equation}
As the volume of the system is a free parameter,
the equilibrium state of the system is obtained by variation over both
the matter field $\Psi$ and the volume $V$:
\begin{equation}
\frac{\delta  E}{\delta \Psi} = 0\,,\quad \frac{d E}{dV}=0~.
\label{eq:Equilibrium2}
\end{equation}
The second equation in  \eqref{eq:Equilibrium2} actually corresponds to the
condition of having no external pressure (cf. Sec.~\ref{sec:external-pressure})
and gives
\begin{equation}
P=-\frac{d E}{dV}=-\epsilon(\Psi_0,q) +q\:\frac{d\epsilon(\Psi_0,q)}{dq}=0~,
\label{eq:Equilibrium3}
\end{equation}
where $\Psi_0$ corresponds to the equilibrium value of the classical matter
field, which has been taken to be spacetime independent.
The solution of \eqref{eq:Equilibrium3} determines the equilibrium value
$q=q_0$ and the corresponding volume is given by $V=V_0=Q/q_0$.

Consider, for the moment, the simplest possible case, where
the energy density $\epsilon(\Psi,q)$
splits into two independent parts, the Ginzburg--Landau functional
for the macroscopic matter field $\Psi(x)$ and the energy density of the
microscopic vacuum $q$--field (both fields taken to be spacetime independent):
\begin{subequations}
\label{eq:epsilonMICROepsilonMACRO}
\beqa
\epsilon(\Psi,q)&=&
\epsilon_\text{micro}(q)
+\epsilon_\text{macro}(\Psi)\,,
\label{eq:epsilonMICROepsilonMACRO-sum}
\\[2mm]
\epsilon_\text{macro}(\Psi)&=&\alpha\, |\Psi|^2 +\beta\, |\Psi|^4~,
\label{eq:epsilonMICROepsilonMACRO-GLpart}
\eeqa
\end{subequations}
for real constants $\alpha$ and $\beta$ with $\beta > 0$.
Now, condition \eqref{eq:Equilibrium3}  becomes
\begin{equation}
P=-\epsilon_\text{macro}(\Psi_0)-\epsilon_\text{micro}(q)
+q\:\frac{d\epsilon_\text{micro}(q)}{dq}=0~.
\label{eq:Equilibrium4}
\end{equation}

For $\alpha>0$ in \eqref{eq:epsilonMICROepsilonMACRO-GLpart},
one has a vanishing equilibrium value of the classical matter field,
$\Psi_0=0$. As a result, the corresponding
energy density vanishes, $\epsilon_\text{macro}(\Psi_0)=0$,
and condition \eqref{eq:Equilibrium4} becomes
\begin{equation}
P=-\epsilon_\text{micro}(q) +q\:\frac{d\epsilon_\text{micro}(q)}{dq}=0~.
\label{eq:EquilibriumPsi=0}
\end{equation}
The solution of this last equation determines the actual equilibrium
value $q=q_0$ of the perfect (nondisturbed) quantum vacuum.
The special case with linear behavior
$\epsilon_\text{micro}(q)\propto q$ needs to be discarded, as
it does not allow the vacuum to reach an equilibrium
(such an energy density would anyway be unbounded from below).

\subsection{Vacuum under external pressure}
\label{sec:external-pressure}

If the portion of quantum vacuum considered is not isolated from
the environment, the pressure of the vacuum in equilibrium equals the
external pressure. Under external pressure $P$, the relevant thermodynamic
potential (Gibbs free energy) at zero temperature is given by
\cite{Pippard1964,LandauLifshitzStatMech1,Perrot1998}
\begin{equation}
W=E+P\,V=\int d^3r~\epsilon\left(\Psi,Q/V\right) + P\,V~,
\label{eq:ThermodynamicPotential}
\end{equation}
with $V$ the volume of the portion of quantum vacuum considered,
$Q\equiv q\,V$ the fixed total ``charge,''  and
$\epsilon\left(\Psi,q\right)$ the energy density introduced in \eqref{eq:P=0}.
The variational equations,
\begin{equation}
\frac{\delta W}{\delta \Psi} = 0\,,\quad \frac{d W}{dV}=0~,
\label{eq:Equilibrium}
\end{equation}
give an integrated form of the Gibbs--Duhem equation:
\begin{equation}
P=-\epsilon(\Psi_0,q) +q\:\frac{d\epsilon(\Psi_0,q)}{dq}~,
\label{eq:Gibbs-Duhem}
\end{equation}
whose solution determines the equilibrium value $q=q_0$
for a definite external pressure $P$ and a fixed ''charge'' $Q\equiv q\,V$.
The Gibbs--Duhem equation \cite{Perrot1998} in its simplest form,
$N\,d\mu= V\, dP - S\, dT$,
relates an infinitesimal change $d\mu$ in the chemical potential $\mu$ to
infinitesimal changes in the other thermodynamic variables of the system,
the change $dP$ in the pressure $P$ and the change $dT$ in the temperature $T$,
with $N$ the conserved particle number and $S$ the entropy.
Here, the system is kept at $T=0$ and the conserved quantity $Q$ plays
the role of the particle number $N$, so that $dE/dQ=d\epsilon/dq$ corresponds
to the chemical potential $\mu$.

It is seen from \eqref{eq:Gibbs-Duhem}
that the thermodynamically relevant vacuum energy density is given by
\begin{equation}\label{eq:widetilde-epsilon-vac}
\widetilde{\epsilon}_\text{vac}(\Psi,q) \equiv
\epsilon(\Psi,q) -q\:\frac{d\epsilon(\Psi,q)}{dq}~,
\end{equation}
for equilibrium values $q=q_0 \equiv q_\text{vac}$ and $\Psi=\Psi_0$.
The numerical values of $q_0$ and $\epsilon(\Psi_0,q_0)$
are determined by the detailed microscopic theory, so that
$\epsilon(\Psi_0,q_0)$ can be expected to be of order
$E^{4}_\text{UV}$, where the ``ultraviolet'' energy scale $E_\text{UV}$ is,
for the moment, taken to be of order $E_\text{Planck}$.
But the vacuum energy density $\widetilde{\epsilon}_\text{vac}$
is determined by macroscopic physics, namely, the external pressure $P$.
It will be argued in Sec.~\ref{sec:NullificationLambda}
that it is the macroscopic vacuum
energy density $\tilde\epsilon_\text{vac}$ which plays the role
of the cosmological
constant rather than the microscopic energy density $\epsilon$.

It is furthermore assumed that the stationary point $(\Psi_0,q_\text{vac})$
of the thermodynamic potential \eqref{eq:ThermodynamicPotential}
corresponds to a minimum, be it local or global.
This implies, in particular, that
\begin{equation}
\left[q^2\;\frac{d^2\epsilon(\Psi_0,q)}{dq^2}\,\right]_{q=q_\text{vac}}
\geq 0~.
\label{eq:Stability}
\end{equation}
With the standard definition \cite{Perrot1998} of the inverse of
the isothermal compressibility, $\chi^{-1} \equiv -V\,dP/dV$, and
the functional behavior $P=P(q)$ from \eqref{eq:Gibbs-Duhem}, it is
possible to identify the left-hand side of \eqref{eq:Stability}
with the inverse vacuum compressibility and to obtain
the following inequality
\begin{equation}
\chi_\text{vac}^{-1} =
\left[q^2\;\frac{d^2\epsilon(\Psi_0,q)}{dq^2}\,\right]_{q=q_\text{vac}}
\geq 0~.
\label{eq:Compressibility}
\end{equation}

\par From the low-energy point of view,
the compressibility of the vacuum $\chi_\text{vac}$ is a fundamental
physical constant, just as Newton's gravitational constant $G$.
The numerical value of $\chi_\text{vac}$
is determined by the detailed microscopic theory and can be
expected to be of order $1/E^{4}_\text{UV}$.
A positive value of the vacuum compressibility
is a necessary condition for the stability of the vacuum in the
absence of an external environment
(cf. Sec.~21 of Ref.~\cite{LandauLifshitzStatMech1}).
It is, in fact, the stability of
the vacuum which leads to the nullification of the cosmological
constant, as will be discussed in Sec.~\ref{sec:NullificationLambda}.

\subsection{Vacuum readjustment}
\label{sec:Readjustment}

The simple model \eqref{eq:epsilonMICROepsilonMACRO} of Sec.~\ref{sec:Variable}
allows us to discuss quantitatively the back reaction
of the low-energy effective field $\Psi(x)$ on the underlying
microscopic subsystem, as described by the thermodynamic parameter $q$
and for the case of zero external pressure.

Consider what happens to the equilibrium value $q_0$
if a cosmological phase transition occurs
in the low-energy effective theory, which is, for example, described
by the electroweak standard model or quantum chromodynamics.
Let us discuss only the simplified case, where the
symmetry-breaking phase transition results from having $\alpha<0$ in
\eqref{eq:epsilonMICROepsilonMACRO-GLpart}. Then, a nonzero value of
$\Psi_0$ results and the Ginzburg--Landau energy density
$\epsilon_\text{macro}(\Psi_0)$ becomes negative.
It follows from condition \eqref{eq:Equilibrium4} that the parameter
$q_0$ is renormalized to $q_0^\prime=q_0+\delta q_0$ after the phase
transition. In words, the parameter value of $q_0$ is readjusted
to the zero value of the pressure in the new  equilibrium state of the vacuum.

Specifically, the relative change of parameter $q_0$ after the transition
is given by
\begin{equation}
\frac{\delta q_0}{q_0}=
\chi_{0}\;\epsilon_\text{macro}(\Psi_0)~,
\label{eq:deltaq}
\end{equation}
where the vacuum compressibility $\chi_{0}$ is defined by the
inverse of the left-hand side of \eqref{eq:Stability} evaluated at
the original equilibrium value $q=q_0$. From \eqref{eq:deltaq}
with $\chi_{0}\sim E^{-4}_\text{UV}$, one finds that a small change of
the microscopic parameter suffices to compensate the Ginzburg--Landau energy
density from a low-energy symmetry-breaking phase transition:
\begin{equation}
|\delta q_0/q_0| \sim \epsilon_\text{macro}/E^4_\text{UV}~.
\label{eq:deltaq2}
\end{equation}
For the case of an electroweak phase transition, the energy density
$\epsilon_\text{macro}$ is determined by the energy scale
$E_\text{ew}\sim M_W \sim 10^2\;\text{GeV}$ and one has
\begin{equation}
|\delta q_0/q_0|\sim E^4_\text{ew}/E^4_\text{UV} \sim 10^{-68}~.
\label{eq:deltaq3}
\end{equation}

Result \eqref{eq:deltaq3} quantifies the smallness of the relative change of
the vacuum parameter for any physical process occurring at
electroweak energies or lower. But the discussion based on
\eqref{eq:Equilibrium4} also suggests that
the renormalization of the fundamental constants in the standard model
from the readjustment of the ``deep-vacuum variable'' $q$ to a
new low-energy state would be negligibly small compared to mechanisms operating
at the low-energy scale itself.

The fine-structure constant $\alpha$, for example,
is determined both by the ultraviolet energy scale
$E_\text{UV}$ and the ``infrared'' electroweak scale $E_\text{ew}$.
Essentially, the inverse of the fine-structure constant is given by
the natural logarithm of the ratio of these two energy scales,
\begin{equation}
1/ \alpha \sim \ln\,( E_\text{UV}/E_\text{ew})~.
\label{eq:alpha}
\end{equation}
The relative change in $\alpha$ due to changes
$\delta E_\text{ew} \sim E_\text{ew}$ of the infrared physics
is of order $\alpha$,  whereas the relative
change in $\alpha$ due to an adjustment of the deep vacuum is given by
\beqa
|\delta \alpha  / \alpha|
&\sim&
|\delta E_\text{UV}/E_\text{UV}|\,\alpha
\sim
|\delta q_0 / q_0|\,\alpha
\sim
(E_\text{ew}^4 / E_\text{UV}^4) \,\alpha
\sim 10^{-68}\,\alpha~.
\label{eq:relativechangealpha}
\eeqa

Anticipating the discussion of gravity in Sec.~\ref{sec:NullificationLambda},
consider also Newton's constant $G$ which
is mainly determined by the ultraviolet scale of the deep vacuum,
\begin{equation}
G^{-1} \sim E_\text{UV}^2 \pm E_\text{ew}^2~.
\label{eq:Ginverse}
\end{equation}
Even though the second term on the right-hand side of \eqref{eq:Ginverse}
is small compared to the first one, it is precisely this term which is
responsible for the main change of $G$.
In fact,  the first term on the right-hand side of \eqref{eq:Ginverse}
gives for the relative change of $G$
due to a change $\delta q_0$ in the equilibrium vacuum parameter $q_0$
the value
$|\delta G / G|^{(1)} \sim |\delta q_0 / q_0| \sim E_\text{ew}^4 / E_\text{UV}^4$,
whereas the second term gives for the relative change of $G$
due to changes $\delta E_\text{ew} \sim E_\text{ew}$
from the low-energy physics
the larger value $|\delta G / G|^{(2)} \sim  E_\text{ew}^2 / E_\text{UV}^2$.
Hence, the total relative  change in $G$ would be of order
\begin{equation}
|\delta G / G| \sim E_\text{ew}^2 / E_\text{UV}^2 \sim 10^{-34}~,
\label{eq:relativechangeG}
\end{equation}
which is still a very small number.

All this implies that the thermodynamics of the vacuum variable $q$
(or, more generally, the dynamics of $q$)
does not influence processes occurring in the low-energy world.
There is, however, one exception, namely,
the cosmological constant which will be discussed in the next section.

\section{Nullification of the cosmological constant $\boldsymbol{\Lambda}$}
\label{sec:NullificationLambda}

\subsection{Effective vacuum energy density}
\label{sec:effective-vacuum-energy-density}

In this section, we address the question of which type of vacuum energy
density would be gravitating in the context of emergent-gravity
models \cite{Bjorken2001,Laughlin2003,FroggattNielsen2005,Volovik2007}.
The local gauge invariance and local Lorentz invariance of our current theories
(standard model and general relativity)
would then be emergent symmetries in the low-energy limit of
some unknown fundamental theory. With general covariance
(i.e., local Lorentz invariance) in place, the effective
low-energy equations of gravitation are the standard Einstein
field equations \cite{Einstein1917,Weinberg1972,Veltman1976},
with Ricci curvature tensor $R^{\mu\nu}(x)$
sourced by the matter energy-momentum tensor $T_\mathrm{M}^{\mu\nu}(x)$
and the cosmological constant $\Lambda$,
\beq\label{eq:classical-field-equations}
\frac{1}{ 8\pi \, G \,}  \Big(R^{\mu\nu}(x) -
\textstyle{\frac{1}{2}} \, g^{\mu\nu}(x)\,
R(x)\Big)= - T_\mathrm{M}^{\mu\nu}(x) - \Lambda\,g^{\mu\nu}(x) \;.
\eeq
Taking the signature of the metric $g_{\mu\nu}(x)$
as \mbox{$(\,+\,-\,-\,-\,)\,$},
the quantity $\Lambda$ in \eqref{eq:classical-field-equations}
corresponds to a vacuum energy density \cite{Bronstein1933,Zeldovich1967}.
As mentioned in the Introduction, the problem is to understand
the small but nonzero value of this  vacuum energy density
(that is, small compared to Planck-scale energy densities).

For the emergent theory envisaged in this article,
the quantum field theory (QFT) of the low-energy fields must satisfy
two requirements:
(1) the resulting QFT must take into account the conservation
of the global quantity $Q$, and
(2) the energy-momentum tensor of the gravitating quantum vacuum must give
the standard relativistic relation between energy and momentum from Lorentz
invariance.

In fact, the general form of the relativistic energy-momentum
tensor of a perfect fluid is given by
\begin{equation}
     T_\text{vac}^{\mu\nu}=\widetilde{\epsilon}_\text{vac}\,u^\mu u^\nu+
P_\text{vac}\,(u^\mu u^\nu-g^{\mu\nu})~,
\label{eq:VacuumEM}
\end{equation}
where  $u^\mu$ is the 4--velocity
of the fluid with respect to the spacetime coordinate system chosen.
For the Lorentz-invariant quantum
vacuum, this tensor must be the same in any coordinate system and cannot
depend on  the 4--velocity  $u^\mu$.
Hence, the equation of state for the vacuum must be
\begin{equation}
P_\text{vac}=-\widetilde{\epsilon}_\text{vac}
\label{eq:VacuumEM1}
\end{equation}
and the vacuum energy-momentum tensor \eqref{eq:VacuumEM} becomes
\begin{equation}
T_\text{vac}^{\mu\nu} =\widetilde{\epsilon}_\text{vac}\,g^{\mu\nu}
     ~.
\label{eq:VacuumEMTensor}
\end{equation}
It is then entirely appropriate to speak of a
``Lorentz-invariant quantum vacuum,'' as a suitable coordinate
transformation brings \eqref{eq:VacuumEMTensor} to the form
$\widetilde{\epsilon}_\text{vac}\,\eta^{\mu\nu}$ in terms of the standard
inverse Minkowski metric $\eta^{\mu\nu}$ $=$ $\text{diag}(1,-1,-1,-1)$.

In order to construct the proper low-energy QFT satisfying the two
requirements mentioned above, we introduce the Hamiltonian
operator corresponding to the simple classical model
\eqref{eq:epsilonMICROepsilonMACRO}:
\begin{equation}\label{eq:QuantumExtension}
    {\cal H}(\widehat{\Psi},q)=
E_\text{micro}(q)
+ {\cal H}_\text{macro}(\widehat{\Psi})~,
\end{equation}
where $\widehat{\Psi}(x)$ is now a quantum field
and $q$ at low energies can be considered as a constant classical parameter
(its quantization will probably be required in the definitive theory
and possibly also its spacetime dependence).
The relevant low-energy Hamiltonian, which takes $Q$--conservation
into account, is then represented by the quantum counterpart of
\eqref{eq:widetilde-epsilon-vac}:
   \begin{equation}
\widetilde{\cal H}=
{\cal H}- q\:\frac{d{\cal H}}{d q}~.
\label{eq:TildeH}
\end{equation}
The vacuum energy experienced by the low-energy degrees of freedom is
the vacuum expectation value of this Hamiltonian:
\beqa
<{\rm vac}|\,\widetilde{\cal H}\,|{\rm vac}>
&=&
\left<{\rm vac}|\, {\cal H}\,|{\rm vac}\right>
-q\:\frac{d}{d q}\left<{\rm vac}\left|\, {\cal H}\,\right|{\rm vac}\right>
 =               
V \epsilon_\text{vac}-V q\:\frac{d\epsilon_\text{vac}(q)}{d q} ~.
\label{eq:vev}
\eeqa
The corresponding vacuum energy density is found to be given by
\begin{equation}
\widetilde{\epsilon}_\text{vac}
=
\epsilon_\text{vac}- q \frac{d\epsilon_\text{vac}}{dq}
=
-P_\text{vac} ~,
\label{eq:VacuumEenergy}
\end{equation}
where the last equality follows from the
Gibbs--Duhem relation \eqref{eq:Gibbs-Duhem} by equating the external
pressure $P$ with the internal pressure $P_\text{vac}$.
The energy density and pressure \eqref{eq:VacuumEenergy}
obtained by a thermodynamic argument are, indeed, seen to satisfy
condition \eqref{eq:VacuumEM1} from Lorentz-invariance.

The Hamiltonian for the low-frequency excitations is
built upon the ground state of \eqref{eq:TildeH}  and reads in
quadratic approximation:
\beqa
\hspace*{-10mm}  
\widetilde{\cal H}
&=&
<{\rm vac}|\,\widetilde{\cal H}\,|{\rm vac}> +
\sum_{n,\,\mathbf{p}}
E^{(n)}_{\bf p}\,a_{\bf p}^{(n)\,\dagger}\; a_{\bf p}^{(n)}
+ \cdots
 =               
\int d^3r~\widetilde{\epsilon}_\text{vac}+
\sum_{n,\,\mathbf{p}}
E^{(n)}_{\bf p}\,a_{\bf p}^{(n)\,\dagger}\; a_{\bf p}^{(n)} + \cdots~,
\nonumber \\&&
\label{eq:quadratic}
\eeqa
with a sum over quasiparticle type $n$
and energy $E_{\bf p}^{(n)} \equiv \sqrt{|{\bf p}|^2 +m_n^2}$
for a quasiparticle of momentum ${\bf p}$ and effective mass $m_n$.
Note that the quadratic term in \eqref{eq:quadratic} includes
the two polarization modes of the massless graviton.

It is then the energy density $\widetilde{\epsilon}_\text{vac}$ from the
vacuum expectation value of the low-energy Hamiltonian \eqref{eq:TildeH},
and not $\epsilon_\text{vac}$ from \eqref{eq:QuantumExtension},
which gravitates. Correspondingly, only the
excitations above $\widetilde{\epsilon}_\text{vac}$ gravitate,
as indicated by \eqref{eq:quadratic}.
The reason is that it is only $\widetilde{\epsilon}_\text{vac}$
which equals the negative of the pressure of the vacuum.
Hence, $\widetilde{\epsilon}_\text{vac}$ is the
vacuum energy density which corresponds
to the cosmological constant $\Lambda$ in the gravitational field
equations \eqref{eq:classical-field-equations}:
\begin{equation}
\Lambda=\widetilde{\epsilon}_\text{vac}(q_0)=-P_\text{vac}(q_0)~,
\label{eq:Lambda}
\end{equation}
for an equilibrium value $q_0$ of the vacuum parameter $q$
(the numerical value of $q_0$ needs to be determined by a further condition
such as pressure equilibrium in Sec.~\ref{sec:Nullification}).
Conceptually, the identification of the
cosmological constant $\Lambda$ with the self-tuned vacuum energy density
$\widetilde{\epsilon}_\text{vac}(q_0)$
may be one of the most interesting results of the present article.

The following two remarks elaborate on the reasoning of the
previous paragraphs. The first remark  concerns precisely
the emergence of the effective cosmological constant
\eqref{eq:Lambda} in our approach to the problem.
In this article, we have discussed certain properties of a
Lorentz-invariant self-sustained quantum vacuum
in which gravity (fundamental or emergent) is already
supposed to exist and have obtained the low-energy Hamiltonian
\eqref{eq:TildeH} which is consistent with this kind of vacuum.
Pressure in the effective theory
must coincide with pressure in the microscopic theory and the same
holds for temperature. The thermodynamic quantities $P$ and $T$
are, therefore, invariant quantities, that is, they do not depend on
the detailed theory but are determined by the environment.
In the low-energy effective theory
with the single remaining microscopic variable $q$,
the equation of state of the vacuum is given by
$P=-<{\rm vac}|\,{\cal H}_\text{eff}\,|{\rm vac}>$, and
this particular equation of state is certainly consistent with gravity,
i.e., with having a $\Lambda$ term in \eqref{eq:classical-field-equations}.
In the fully microscopic theory with variable $q$ included,
the Gibbs--Duhem relation \eqref{eq:Gibbs-Duhem} reads
$P=-\epsilon_\text{micro}(q) +q\,(d\epsilon_\text{micro}/dq)$.
Since these two pressures must be equal
(that is, the one from the microscopic theory and the one from
the effective theory), we must choose for the relevant effective
Hamiltonian ${\cal H}_\text{eff}$ of the low-energy theory with gravity
the Hamiltonian $\widetilde{\cal H}$ from \eqref{eq:TildeH}
and not ${\cal H}$ from \eqref{eq:QuantumExtension}.
It is, therefore, the vacuum expectation value of $\widetilde{\cal H}$
which gravitates and not the vacuum expectation value of ${\cal H}$.
However, it must be admitted that the argument just given remains
heuristic in the absence of the detailed microscopic theory.
Still, a dynamic origin of $\widetilde{\epsilon}_\text{vac}(q)$
is, in principle, possible, as will be shown
by the examples of Secs.~\ref{sec:Possible-dynamic-origin-4form}
and \ref{sec:Possible-spacetime-origin-aether}.

The second remark concerns the way how emergent gravity couples to
the low-energy degrees of freedom. The crucial observation, here, is that
emergent (or fundamental) Lorentz gauge invariance is known
\cite{Weinberg1964} to require the equivalence principle for consistency.
The equivalence principle, in turn, underlies Einstein's theory of
gravitation, whose field equations \eqref{eq:classical-field-equations}
were already used in the first paragraph of this subsection.
See also Sec.~10.8 of Ref.~\cite{Weinberg1972} and Sec.~13 of
Ref.~\cite{Veltman1976} for a general discussion of the connection
between local Lorentz gauge invariance and the equivalence principle.

\subsection{Nullification of $\boldsymbol{\Lambda}$ in a perfect
quantum vacuum}
\label{sec:Nullification}

For the perfectly homogenous equilibrium quantum vacuum,
one has in the absence of an environment
(i.e., in the absence of external pressure $P$) a vanishing
cosmological constant from \eqref{eq:Lambda}:
\begin{equation}
\Lambda =\widetilde{\epsilon}_\text{vac}(q_0) =-P_\text{vac}(q_0)=-P=0~.
\label{eq:Lambda-zero}
\end{equation}
Observe that the vanishing of the effective vacuum energy
density $\widetilde{\epsilon}_\text{vac}$ for $q=q_0$
results from the cancelation of the two terms
in the middle part of \eqref{eq:VacuumEenergy},
each of which are typically of order $E_\text{UV}^4$.
Numerically, the nullification of $\widetilde{\epsilon}_\text{vac}$
(and, thereby, of $\Lambda$) may be one of the most important results
of the present article and a concrete example will be given in the
last paragraph of Sec.~\ref{sec:Possible-dynamic-origin-4form}.

This zero value for the effective energy density of the isolated quantum
vacuum is consistent with the Lorentz invariance of a perfect vacuum.
Indeed, the mixed terms $T^i_0$ in \eqref{eq:VacuumEMTensor}
vanish for the inverse Minkowski metric $\eta^{\mu\nu}$
and the 3--momentum of the vacuum is zero, ${\bf P}=0$. The
relativistic relation \cite{Einstein1907}
between the system velocity ${\bf v}$, momentum ${\bf P}$, and energy
$\widetilde{E}$
is given by (temporarily reinstating $c$)
\begin{equation}
c\,{\bf P} = \widetilde{E} \;\frac{{\bf v}}{c}~.
\label{eq:LorentzEnergyMomentum}
\end{equation}
Having ${\bf P}=0$ then requires that the vacuum energy also vanishes,
$\widetilde{E}\equiv \int d^3r ~\widetilde{\epsilon}_\text{vac}= 0$.
Physically, this result can be understood as follows.
If $\widetilde{E}$ would be nonzero,
it would transform under a Lorentz boost and produce a nonzero momentum of
the vacuum. This would imply the existence of a preferred reference frame
in the vacuum, disagreeing with experimental fact \cite{Einstein1907}.
But this disagreement does not arise if $\widetilde{E}$ vanishes, as
the perfect equilibrium quantum vacuum
with $|{\bf P}|=\widetilde{E}=0$ does not have a preferred frame.

A preferred frame does appear if the quantum vacuum is perturbed by
external pressure. The experimentally observed Lorentz invariance
(cf. Refs.~\cite{KosteleckyMewes2002,KlinkhamerRisse2007} and references
therein) suggests that there is no external pressure and that the vacuum
energy density must be zero. Lorentz invariance may still be violated by
the existence of spacetime boundaries or the presence of matter,
both of which introduce preferred reference frames.
(With matter present, the preferred reference frame is the one
in which the matter is, on average, at rest.)
In all these cases, the vacuum energy density becomes nonzero
but small compared to a Planck-scale energy density, as will be
discussed further in the next section.

The violation of Lorentz invariance by the presence of matter, spacetime
boundaries, or external pressure does not mean that Lorentz invariance,
as a \emph{law of physics},
is violated. Instead, it is the \emph{state} of the universe which looses
the property of Lorentz invariance
in the presence of matter and/or external environment,
precisely because of the appearance of a preferred frame. Throughout this
article, we assume that Lorentz invariance is either a fundamental physical
law or a symmetry which can only be violated at an energy scale $E_\text{LV}$
far above the Planck energy scale (cf. Ref.~\cite{KlinkhamerVolovik2005JETPL}).

\section{Nonzero vacuum energy density from perturbations}
\label{sec:Nonzero-vacuum-energy-density}

\subsection{Preliminary remarks}
\label{sec:Preliminaries}

In sec.~\ref{sec:Readjustment}, we have discussed the response
of the ``deep vacuum'' to a homogeneous change of the matter ground
state by a low-energy phase transition, which is, for simplicity,
described in terms of the
Ginzburg--Landau functional of a single matter field.
In this section, we consider the response of the deep vacuum to
perturbations which themselves violate Lorentz invariance.
The outstanding problem is to see how this response
can be consistent with Lorentz invariance as a physical law.
In Sec.~\ref{sec:external-pressure},
we have already discussed this issue by using the example of an external
pressure which, indeed, violates Lorentz invariance.
But external pressure is, by definition, not accessible to us
and, here, we wish to consider other physically more
relevant perturbations of the perfect quantum vacuum.

Two types of Lorentz-noninvariant perturbations will be discussed in this
section, the presence of matter (e.g., homogeneous thermal matter)
and the possibility of having a nonuniform setup
(e.g., a droplet with a different content
than the ambient space). In both cases, we will investigate how
and under which conditions a Lorentz-invariant (relativistic) relation
between energy and momentum arises.

\subsection{Vacuum in the presence of matter}
\label{sec:matter}

As a first \emph{Gedankenexperiment}, consider thermal
matter enclosed in a box. This box is taken to have
rigid walls which are impenetrable for the matter contained in the box
and the box as a whole is considered to be moving in an
empty spacetime (perfect quantum vacuum)
with the Minkowski metric in standard coordinates. Furthermore,
assume that the box is, on the one hand, sufficiently small so as
to ignore its gravitational effects and, on the other hand, sufficiently
large in size so as to ignore the energy of the walls of the box
compared to the energy of the matter inside the box.

Let  $v^\mu$  be  the 4--velocity of the box.
The energy-momentum tensor of the matter inside the box is
\begin{equation}
     T_\mathrm{M}^{\mu\nu}=\rho_\text{M}\,v^\mu v^\nu+
P_\text{M}\,(v^\mu v^\nu-g^{\mu\nu})
     ~,
\label{eq:MatterEM}
\end{equation}
where $\rho_\text{M}$ and $P_\text{M}$ are
the energy  density and pressure
of matter in the frame moving along with the box. A naive consideration
would suggest that the energy $E$ and momentum $P_i$
of the box with matter
are given by integration of, respectively,
$T_\mathrm{M}^{00}$ and $T_\mathrm{M}^{0i}$
over the box volume. But this cannot be correct,
since the resulting energy $E$ and momentum $P_i$
would not satisfy the relativistic relation \eqref{eq:LorentzEnergyMomentum}.
In fact, one must also take into account the work needed to compress the matter
into the box. For the correct energy-momentum tensor component $T^{00}$,
one must therefore use $\rho_\text{M}+P_\text{M}$
instead of $\rho_\text{M}$ in \eqref{eq:MatterEM}.

However, there is a physical way to avoid wall pressure, namely,
by having a phase boundary (interface) separating two states
which are in equilibrium with each other. Specifically, consider
having a perfect quantum  vacuum outside the interface and
a vacuum with matter inside the interface.
Then, the outer vacuum has zero pressure and the
partial pressure of the inner vacuum compensates the partial
pressure of matter, as long as the surface tension can be neglected.
The pressure is now zero both inside and outside of the box, so that an
area element of the box wall (interface) experiences no net force.

The total energy-momentum tensor of the system inside the box then
has two contributions:
\beqa
T^{\mu\nu}
&\equiv&
T_\mathrm{M}^{\mu\nu}+ T_\text{vac}^{\mu\nu}
 =               
(\rho_\text{M}+\widetilde{\epsilon}_\text{vac})\,v^\mu v^\nu +
(P_\text{M}+P_\text{vac})(v^\mu v^\nu-g^{\mu\nu})~.
\label{eq:Matter+Vac}
\eeqa
If the vacuum pressure compensates the matter pressure,
$P_\text{M}+P_\text{vac}=0$, one obtains
\begin{equation}
     T^{\mu\nu}  =
     (\rho_\text{M}+\widetilde{\epsilon}_\text{vac})\,v^\mu v^\nu ~,
\label{eq:Box-Tmunu}
\end{equation}
with
\begin{equation}
\widetilde{\epsilon}_\text{vac}=-P_\text{vac}=P_\text{M}~.
\label{eq:Box-widetilde-epsilon-vac}
\end{equation}
The integration of $T^{00}$ and $T^{0i}$  over the box volume now gives the
correct relativistic relation \eqref{eq:LorentzEnergyMomentum} between the
energy and momentum of the box. It is, therefore, a genuine relativistic
object whose rest
mass is given by $(\rho_\text{M}+\widetilde{\epsilon}_\text{vac})V_0$,
where $V_0$ is the volume of the box in a comoving frame. This effective
rest mass is, in fact, the sum of matter energy and vacuum energy.
But, as mentioned above, the box must be relatively small so as to
ignore all gravitational effects.

It is also possible to have an interface between \emph{vacua}
with different energy
densities, namely, by considering a cosmic domain wall separating two different
types of vacua (for example, ``false'' and ``true''  vacua,
as will be explained later on).
Such an interface has a large energy itself and we shall take
this surface energy into account in Sec.~\ref{sec:droplet}.

\subsection{Universe with nongravitating matter}
\label{sec:nongravitating-matter}

As a further \emph{Gedankenexperiment}, expand the box wall of the
previous subsection to a shell enclosing the entire visible universe,
with the  shell corresponding to what might be called the ``boundary''
of the universe.
This droplet universe \cite{Volovik2003,KlinkhamerVolovik2005PLA},
containing both vacuum and matter,
would again be moving in empty spacetime
(perfect quantum vacuum) with the Minkowski metric in standard coordinates.

In this article, we are primarily interested in the properties of the
quantum vacuum which are governed by Lorentz invariance.
The presence of gravitating matter, however, leads to a
deviation of the metric from the Minkowski metric and  Lorentz invariance
looses its meaning. That is why, here, we will only consider the
hypothetical case of a universe filled with ``nongravitating''
matter \cite{Volovik2003}, where Newton's constant $G$ is effectively set to zero.
In such a  hypothetical universe, the metric $g_{\mu\nu}$ remains constant and
takes the Minkowski form in appropriate coordinates.

In a general coordinate frame, the energy and momentum density
of matter are given by \cite{Weinberg1972,Einstein1907}
\begin{subequations}
\label{eq:MatterEnergyMomentum}
\beqa
    \rho_\text{M}(v)
    &=&
     \frac{\rho_\text{M} +(v^2/c^2)P_\text{M}}{1- v^2/c^2} \,,
\label{eq:MatterEnergyMomentum-rho}
\\[2mm]
     {\bf p}_\text{M}(v)
     &=&
     \frac{{\bf v}}{c^2}~
     \frac{\rho_\text{M} +P_\text{M}}{1- v^2/c^2}~,
\label{eq:MatterEnergyMomentum-p}
\eeqa
\end{subequations}
with $c$ reinstated for clarity.
One obvious consequence of \eqref{eq:MatterEnergyMomentum} is
that energy and momentum of matter do not satisfy the
relativistic relation \eqref{eq:LorentzEnergyMomentum}.
As mentioned in Sec.~\ref{sec:matter},
the reason for this has to do with the presence of external forces acting on
matter, which violate Lorentz invariance
(these forces establish a preferred reference frame in which they are
isotropic).
The external forces manifest themselves in
\eqref{eq:MatterEnergyMomentum} through the pressure $P_\text{M}$ of the matter,
which is supported by the external pressure $P$
(see, e.g., Part IV  of Ref. \cite{Einstein1907}). If the universe would be
completely isolated from the ``environment,'' the external pressure
would be absent, $P=P_\text{M}=0$, and the Lorentz-invariant
equation \eqref{eq:LorentzEnergyMomentum} would be restored.

But the typical matter considered in realistic cosmological models
\cite{Weinberg1972},
such as a  relativistic plasma, cannot exist as an equilibrium state
at zero pressure (except for the case of cold dust with $P_\text{M}=0$).
According to special relativity, the equilibrium universe would appear
to be necessarily empty of matter.
However, in this consideration, we have not yet taken into account the
contribution of the quantum vacuum with equation of state \eqref{eq:VacuumEM1}.

The equilibrium state of the system (quantum vacuum + nongravitating matter)
is achieved when the partial pressure of matter is compensated by
the partial pressure of the vacuum.
In that case, the required external pressure $P$ can be zero,
\begin{equation}
P=P_\text{vac}+P_\text{M}=0 ~,
\label{eq:VacuumEM2} \end{equation}
and the system can be in equilibrium without external
environment.  For this equilibrium universe, the second term in
\eqref{eq:Matter+Vac} disappears and the relativistic relation
\eqref{eq:LorentzEnergyMomentum} between the energy and momentum of the
whole universe (droplet) is restored:
\begin{equation}
     T^{\mu\nu}\equiv T_\text{M}^{\mu\nu}+ T_\text{vac}^{\mu\nu}=
     (\rho_\text{M}+\widetilde{\epsilon}_\text{vac})\,v^\mu v^\nu
     .
\label{eq:Matter+VacEquil}
\end{equation}

Using the equation of state \eqref{eq:VacuumEM1} and the equilibrium
condition \eqref{eq:VacuumEM2}, one obtains
the effective vacuum energy density
induced by nongravitating matter with pressure $P_\text{M}\,$:
\begin{equation}
\widetilde{\epsilon}_\text{vac}=-P_\text{vac}=P_\text{M}~,
\label{eq:VacuumEM3}
\end{equation}
which is of the same form as \eqref{eq:Box-widetilde-epsilon-vac}
but now for a droplet universe with fictitious (nongravitating) matter.
Since the vacuum momentum density is zero,
\begin{equation}
{\bf p}_\text{vac}=0~,
\label{eq:VacuumMomentum}
\end{equation}
the total energy and momentum densities of the system
(quantum vacuum + nongravitating matter) become
\begin{subequations}
\label{eq:TotalEnergyMomentum}
\beqa
\rho_\text{total}&=&
\rho_\text{M}(v)+\widetilde{\epsilon}_\text{vac}=
\frac{\rho_\text{M} +\widetilde{\epsilon}_\text{vac}}{1- v^2/c^2} \,,
\label{eq:TotalEnergyMomentum-rho}
\\[2mm]
{\bf p}_\text{total}&=&
{\bf p}_\text{M}(v)
+{\bf p}_\text{vac}=  \frac{{\bf v}}{ c^2}~\frac{\rho_\text{M} +
\widetilde{\epsilon}_\text{vac}}{ 1-v^2/c^2}~,
\label{eq:TotalEnergyMomentum-p}
\eeqa
\end{subequations}
where \eqref{eq:MatterEnergyMomentum} and \eqref{eq:VacuumEM3}
have been used. These quantities are seen to satisfy the relativistic
equation \eqref{eq:LorentzEnergyMomentum} and
the corresponding rest-mass density $\mu_\text{rest}$ is given by
\begin{subequations}
\label{eq:TotalRestMass}
\beqa
\rho_\text{total} &=&
\frac{\mu_\text{rest}\,c^2}{ \sqrt{1- v^2/c^2}} \,,
\label{eq:TotalRestMass-rho}
\\[2mm]
\mu_\text{rest}&=&
\frac{\rho_\text{M} +\widetilde{\epsilon}_\text{vac}}{ c^2\sqrt{1- v^2/c^2}} ~,
\label{eq:TotalRestMass-mu}
\eeqa
\end{subequations}
where $c$ is again shown explicitly.

For the calculation of the total rest mass of the droplet,
the extra factor $\sqrt{1-  v^2/c^2}$ in the denominator of
the rest-mass density $\mu_\text{rest}$ is canceled by the relativistic
transformation of the volume. Indeed, the volume element $dV$
in the frame of measurement and the volume element
$dV_0$ in the comoving frame are related by
$dV=dV_0\sqrt{1-  v^2/c^2}\,$, so that the total rest
energy of the system is given by
\beqa
M_\text{rest}\,c^2
&=&
\int d V\, \mu_\text{rest}\,c^2
=\int d V\, \frac{\rho_\text{M} +\widetilde{\epsilon}_\text{vac}}{\sqrt{1- v^2/c^2}}
 =               
(\rho_\text{M} +\widetilde{\epsilon}_\text{vac})\,V_0~.
\label{eq:TotalRestEnergy}
\eeqa
Even though each of the two subsystems (vacuum or matter)
does not separately satisfy the relativistic
relation \eqref{eq:LorentzEnergyMomentum} between energy and momentum,
the whole droplet universe  (vacuum + matter)
does represent a genuine relativistic object whose effective
rest mass is the sum of the energies of matter and quantum vacuum.
This universe has a preferred reference frame which is precisely
the rest frame of the matter.
The state of the universe is no longer Lorentz invariant and this
allows for having a nonzero value of the vacuum energy,
as given by \eqref{eq:VacuumEM3}.

If the matter pressure vanishes, the connection between vacuum and
preferred reference frame of matter is lost and the  zero value
of the Lorentz-invariant vacuum energy restored.
Still, the general result \eqref{eq:VacuumEM3} is of considerable
interest as it relates a relatively small (i.e., sub-Planckian),
and positive, effective vacuum energy density to the presence of
thermal matter, even though the matter considered here
was ``nongravitating.'' The approach of the present article
(suitably extended to genuine gravitating matter)
may provide a new point of view on the
so-called ``cosmic-coincidence'' puzzle of current cosmological
models \cite{Zlatev-etal1998,Padmanabhan2002}.

\subsection{Droplet of false vacuum}
\label{sec:droplet}

The final \emph{Gedankenexperiment} of this section
takes into account possible surface-tension effects of the
droplet universe. We consider a spherical droplet of false vacuum
moving in true vacuum. We assume that the external
true vacuum is in equilibrium, so that, with zero external pressure,
its pressure and energy density vanish.
The vacuum inside the droplet is taken to be ``false,''
meaning that it corresponds to a local minimum
of the Ginzburg--Landau potential
which is higher than the global minimum of the ``true'' vacuum.
The false vacuum has, therefore, a nonzero positive energy density.
(If the false vacuum were on the outside, it would
have vanishing pressure and energy density due to the zero external
pressure, whereas the true vacuum on the inside would have negative energy
density.) We also assume that matter is massive in the true vacuum
but massless in the false vacuum, so that this type of matter cannot pass
through the interface. Hence, the matter is trapped inside the droplet.

Now, the interface between the vacua participates in the dynamics,
along with the trapped matter and vacuum energy.
This interface can be represented by fictitious matter which is distributed
homogeneously over the interior of the droplet and
has the following equation of state:
\begin{equation}
P_\sigma =-\frac{2}{3}\,\rho_\sigma=-\frac{2\sigma}{R}~,
\label{eq:TensionEnergyMomentum}
\end{equation}
where $\sigma$ is the surface tension and $R$  the radius
of the spherical droplet. Expression \eqref{eq:TensionEnergyMomentum}
follows from the free energy $F = A\,\sigma$
and the identification $\rho_\sigma\equiv F/V$
for a spherical volume $V=(4\pi/3) R^3$ with surface area $A=4\pi R^2$;
cf. Ref.~\cite{Pippard1964}. As the interface with nonzero (positive) surface
tension prefers to shrink, this fictitious matter has negative pressure.

If ${\bf v}$ is the 3--velocity of the droplet with
matter energy and momentum densities given by \eqref{eq:MatterEnergyMomentum},
the total droplet energy and momentum per unit volume are given by
\begin{subequations}
\label{eq:DropletEnergyMomentum}
\begin{eqnarray}
     \rho_\text{total}&=&
     \frac{\rho_\sigma+(v^2/c^2) P_\sigma+\rho_\text{M}+(v^2/c^2)P_\text{M}}
          {1- v^2/c^2}
 +               
\widetilde{\epsilon}_\text{vac} ~
\label{eq:DropletEnergyMomentumRho}
\\[2mm]
{\bf p}_\text{total}&=&
\frac{\bf v}{ c^2}\;
\frac{\rho_\sigma+P_\sigma+\rho_\text{M}+P_\text{M}}{1- v^2/c^2}~,
\label{eq:DropletEnergyMomentumP}
\end{eqnarray}
\end{subequations}
with positive energy density $\widetilde{\epsilon}_\text{vac}$
of the interior (false) vacuum and vanishing vacuum  momentum density
according to \eqref{eq:VacuumMomentum}.
In equilibrium, one has pressure balance, so that
\cite{KlinkhamerVolovik2005PLA}
\begin{equation}
\widetilde{\epsilon}_\text{vac}=- P_\text{vac}=P_\sigma+P_\text{M}~.
\label{eq:DropletEnergydensity}
\end{equation}
Observe that, in the absence of a false vacuum
($\widetilde{\epsilon}_\text{vac}=0$), the matter pressure $P_\text{M}$
is given by the Laplace pressure $2\sigma/R$
of the droplet \cite{Perrot1998}.

Inserting expression \eqref{eq:DropletEnergydensity}
for $P_\sigma+P_\text{M}$ into
\eqref{eq:DropletEnergyMomentum}, one obtains that the whole object
(nongravitating matter + false vacuum + interface)
obeys the relativistic relation \eqref{eq:LorentzEnergyMomentum}
between energy and momentum:
\begin{equation}
      {\bf p}_\text{total} = \rho_\text{total}\; \frac{{\bf v}}{c^2}
      =\frac{\rho_\sigma +\rho_\text{M}+\widetilde{\epsilon}_\text{vac}}
            {1-v^2/c^2}\; \frac{{\bf v}}{c^2}~,
\label{eq:DropletEnergyMomentum2}
\end{equation}
whereas the subsystems do not separately obey the relativistic relation.
Result \eqref{eq:DropletEnergyMomentum2} provides a concrete
example of some of the ingredients which may enter a realistic
description of the vacuum energy density indicated by recent
astronomical results \cite{deBernardis2000,Hinshaw2006,Riess2006}.

\section{Nature and origin of the vacuum variable $\boldsymbol{q}$}
\label{sec:Nature-and-origin}

\subsection{Examples of spacetime-independent variables}
\label{sec:Spacetime-independent-variables}

In view of its origin, the proposed parameter $q$ of
the Lorentz-invariant self-sustained quantum vacuum
may be strictly spacetime independent. There are known physical
systems where such constant variables occur as extra degrees of freedom.
In the thermodynamic limit, one then has to minimize over these variables,
which leads to the phenomenon of self-tuning \cite{PolyakovPrivateComm}.

Let us give three examples of these constant variables.
First, there is the homogeneous deformation $u_m({\bf x})=u_{mn}\,x_n$
in solids, where all six components of the strain tensor
$u_{mn}$ are independent, even though, for the nonuniform case
(momentum ${\bf k}\ne 0$), only three components of the
displacement field $u_m$ determine the strain tensor by
$u_{mn}\sim i\,k_m\,u_n + i\,k_n\,u_m\,$.
There is thus a jump in the number of degrees of freedom
at ${\bf k}=0$ compared to ${\bf k}\neq 0$.
The macroscopic elastic mode with ${\bf k}=0$ plays, in fact,  a
crucial role in the magnetic phase transitions of crystals:
the coupling of this mode to the order parameter transforms
the second-order phase transition to a first-order one
(the Larkin--Pikin effect \cite{LarkinPikin1969}).

Second, there can, in principle, exist constant electric and magnetic
fields \emph{in vacuo}, with linear gauge potentials
$A_\mu(x)=-\half F_{\mu\nu}\,x^\nu$ for constant $F_{\mu\nu}=-F_{\nu\mu}$.
These constant electric and magnetic fields
would be mutually independent, in contrast to the spacetime-dependent
fields connected by the Maxwell equations.

Third, there exist so-called ``discrete states'' in certain string theories,
which correspond to homogeneous degrees of freedom of
the system \cite{Polyakov1991}.

\subsection{Possible dynamic origin of $\boldsymbol{q}$: General considerations}
\label{sec:Possible-dynamic-origin-general}

In Sec.~\ref{sec:Variable}, the basic vacuum variable $q^{\mu\nu}$,
with vanishing covariant divergence \eqref{eq:conservation},
was simply postulated.  But what kind of variable can this be?
One possible scenario is the following.

In our approach, we have focussed on only one thermodynamic degree of
freedom out of the large number of degrees of freedom contained
in the quantum vacuum.
At the microscopic level, all degrees of freedom (including the
low-energy effective ones) are described by some
type of underlying QFT, which can be a continuous or discrete theory.
This microscopic QFT  has its own microscopic energy-momentum tensor
$(t_\text{micro})^{\mu\nu}\,$,
which, in principle, has no direct
relation to the energy-momentum tensor of the low-energy effective theory.
For this reason, the tensor $q^{\mu\nu}$ may very well correspond to the
vacuum expectation value of the energy-momentum tensor of the microscopic
quantum fields:
\begin{subequations}
\label{eq:VacuumMatterField}
\begin{eqnarray}
q^{\mu\nu}
&=&
\left<{\rm vac}\left|\,(t_\text{micro})^{\mu\nu}\,\right|{\rm vac}\right>
=q\,g^{\mu\nu}~,
\label{eq:VacuumMatterField-qmunu}\\[2mm]
q
&=&
\textstyle{\frac{1}{4}}\,\left<{\rm vac}\left|\,(t_\text{micro})^{\mu}_{\phantom{\mu}\mu}\,
\right|{\rm vac}\right> .
\label{eq:VacuumMatterField-q}
\end{eqnarray}
\end{subequations}
The energy-momentum conservation law
is assumed to be exact at the microscopic level, so that
$q^{\mu\nu}$, defined by \eqref{eq:VacuumMatterField-qmunu} with constant $q$,
automatically satisfies conservation law \eqref{eq:conservation}.
However, the energy-momentum conservation law for the emergent effective
quantum fields is only approximate, as exchange of energy and momentum is
possible between matter and quantum vacuum.
In other words, energy-momentum can be exchanged
between the low-energy degrees of freedom
visible to the ``poor physicist'' of Ref.~\cite{FroggattNielsen2005}
and the high-energy degrees of freedom from the deep vacuum.

In this article, we have considered the low-energy effective theory
supplemented by a single thermodynamic parameter $q$,
and calculated in Sec.~\ref{sec:Readjustment} the back reaction of
the low-energy effective fields on this parameter $q$,
after they have reached a new equilibrium.
The challenge is to describe the complete process of
equilibration, which allows us to discuss the dynamics of $q$ and, hence,
the dynamics of the ``cosmological constant'' \cite{Barcelo2007};
see Sec.~\ref{sec:Summary} for further remarks.

According to \eqref{eq:VacuumMatterField}, the vacuum variable
$q$ would be determined by the ultraviolet cut-off, $q\sim E_\text{UV}^4$,
where the ultraviolet energy scale $E_\text{UV}$
can be approximately equal to the energy scale  $E_\text{Planck}$
defined in Sec.~\ref{sec:Variable} or a Lorentz-violating energy scale
$E_\text{LV}\gg E_\text{Planck}$ \cite{KlinkhamerVolovik2005JETPL}.
These energy scales are, in fact, the natural
energy scales for any quantity describing the quantum vacuum, but not for
thermodynamic variables (such as pressure and temperature) which are
determined by external conditions and do not depend on the details of
the microscopic physics.

\subsection{Possible dynamic origin of $\boldsymbol{q}$: Four-form field strength}
\label{sec:Possible-dynamic-origin-4form}

A specific suggestion for a possible dynamic origin of
the vacuum variable $q$ starts from a rank-three antisymmetric tensor
gauge potential $A_{\mu\nu\rho}$, with field strength
\cite{DuffNieuwenhuizen1980,Aurilia-etal1980}
\begin{equation}
F_{\mu\nu\rho\sigma}\equiv\nabla_{[\sigma}\,A_{\mu\nu\rho]}\,,
\label{eq:rank4}
\end{equation}
where the square brackets around the spacetime indices denote total
anti-symmetrization.
This gauge potential $A_{\mu\nu\rho}$ may correspond to
a microscopic degree of freedom from the quantum vacuum, as discussed in
the previous subsection.

The classical action of this four-form field strength coupled to gravity
is given by
\begin{equation}
S= S_\text{grav} +S_\text{M}= \int_{\mathbb{R}^4}\, d^4x \,\sqrt{-g}\,
   \left(\frac{R}{16\pi G}   - \epsilon(F)\right)\,,
\label{eq:action}
\end{equation}
with $R$ the Ricci scalar of the pure gravity action $S_\text{grav}$ and
$\epsilon(F)$ a scalar function entering the matter action $S_\text{M}$
with the following definition for the square of its argument:
\begin{equation}
F^2 \equiv -  \frac{1}{24}\,
F_{\mu\nu\rho\sigma}\, F^{\mu\nu\rho\sigma}\,,
\label{eq:Fdefinition}
\end{equation}
where the right-hand side contains four factors of the inverse metric
$g^{\kappa\lambda}$ to convert one covariant tensor
\eqref{eq:rank4} into a contravariant tensor.
A quadratic function $\epsilon(F)=\frac{1}{2}F^2$
for the matter part of action \eqref{eq:action} is often used
in the literature (see, e.g.,
Refs.~\cite{DuffNieuwenhuizen1980,Aurilia-etal1980,Aurilia-etal2004}),
but, here, we keep the even function $\epsilon(F)$ arbitrary.
In general, this function $\epsilon(F)$ involves an
energy scale $E_\text{UV}$ and one has explicitly
$\epsilon(F)=\widehat{\epsilon}(F/E_\text{UV}^2)\,E_\text{UV}^4$
in terms of a dimensionless even function
$\widehat{\epsilon}(f)$ of a dimensionless variable $f$.

The generalized Maxwell equations from \eqref{eq:action} read
\begin{equation}
\nabla^\mu \left(  \frac{1}{F} \frac{d\epsilon(F)}{dF}\,
                  F_{\mu\nu\rho\sigma}\right)=0\,.
\label{eq:Maxwell}
\end{equation}
For a flat spacetime and a Lorentzian metric signature
\mbox{$(\,+\,-\,-\,-\,)\,$},
these equations have the following solution (indicated by a bar):
\begin{equation}
  \overline{F}_{\mu\nu\rho\sigma}=F \,e_{\mu\nu\rho\sigma} \,,\quad F=q\,,
\label{eq:solution}
\end{equation}
with a constant $q$ (given explicitly by $q=\widehat{q}\:E_\text{UV}^2$,
in terms of a dimensionless number $\widehat{q}\,$)
and the totally antisymmetric Levi--Civita symbol $e_{\mu\nu\rho\sigma}$
in a slightly unusual notation (i.e., having the Latin letter `$e$' instead of
the Greek letter `$\epsilon$' which is reserved for the energy density).
For the special case of a quadratic function $\epsilon(F)$,
this solution has already been found by the authors of
Refs.~\cite{DuffNieuwenhuizen1980,Aurilia-etal1980}.

Solution \eqref{eq:solution} gives a constant
(i.e., spacetime-independent) value $\epsilon(q)$
for the energy density which enters action \eqref{eq:action}.
However, the energy-momentum tensor obtained by variation over $g^{\mu\nu}$
in the matter part of action \eqref{eq:action}
is given by
\begin{equation}
T_{\mu\nu}=\frac{2}{\sqrt{-g}}\: \frac{\delta S_\text{M}}{\delta g^{\mu\nu}}=
\epsilon(F)\,  g_{\mu\nu} + \frac{1}{6F}\,  \frac{d\epsilon(F)}{dF}\,
F_{\mu\alpha\beta\gamma}\,  F_\nu^{\phantom{\nu}\alpha\beta\gamma} \,,
\label{eq:em}
\end{equation}
and the following expression results when evaluated for the particular
solution \eqref{eq:solution}:
\begin{equation}
\overline{T}_{\mu\nu}
=T_{\mu\nu}\,
\Big|_{F_{\mu\nu\rho\sigma}=\overline{F}_{\mu\nu\rho\sigma}\,,\;g_{\mu\nu}=\overline{g}_{\mu\nu}}
=\;\overline{g}_{\mu\nu}\,\left[\epsilon(F)-F\,\frac{d\epsilon(F)}{dF}\right]_{F=q}\,,
\label{eq:emSolution}
\end{equation}
where $\overline{g}_{\mu\nu}$ stands for the flat spacetime metric.
Result \eqref{eq:emSolution} corresponds to having the following cosmological
constant in the gravitational field equation \eqref{eq:classical-field-equations}:
\begin{equation}
\Lambda =      \epsilon(q) - q\, \frac{d\epsilon}{dq}
        \equiv \widetilde{\epsilon}_\text{vac}(q)\,,
\label{eq:CC}
\end{equation}
which has precisely the form discussed
in Sec.~\ref{sec:effective-vacuum-energy-density}
(see, e.g., Ref.~\cite{Hawking1984+Duff+Wu} for a related but different
discussion of $\Lambda$ and four-form fields).

Two remarks are in order. First, the transmutation
of the energy density $\epsilon(q)$ of the action
to the effective vacuum energy density $\widetilde{\epsilon}_\text{vac}(q)$
of the gravitational field equations occurs naturally.
The reason is that the quantity $d\epsilon/dF$ is constant according
to \eqref{eq:Maxwell} and \eqref{eq:solution},
so that $d\epsilon/dF$ indeed plays the role of a chemical
potential; see the discussion on the Gibbs--Duhem equation
in Sec.~\ref{sec:external-pressure}.
For a quadratic function $\epsilon(q)$, the sign of
$\widetilde{\epsilon}_\text{vac}(q)$ from \eqref{eq:CC}
is simply the opposite of $\epsilon(q)$,
as noted already in Ref.~\cite[(b)]{Hawking1984+Duff+Wu}.

Second, it is possible to have a nonzero equilibrium value $q_0$
in the vacuum at vanishing external pressure,
$P=0$, for appropriate functions $\epsilon(F)$ in the
original action \eqref{eq:action}.
The equilibrium value $q_0$ is determined
by condition \eqref{eq:EquilibriumPsi=0}, which corresponds to
the nullification of cosmological constant \eqref{eq:CC}.
For example,
the function $\epsilon(F)=\epsilon_\text{b}+\half\,\sin^2 F$,
with the energy scale $E_\text{UV}$ set to unity and
the ``bare'' cosmological constant $\epsilon_\text{b}$
set to the arbitrary numerical value $1/5$,
gives a possible equilibrium value $q_0 \approx 3.20479$
[from the solution of the zero-pressure Gibbs--Duhem relation
 $(1/5)+(1/2)\, \sin^2 q - q\, \sin q \,\cos q=0\,$]
and the corresponding inverse vacuum compressibility
$\chi_0^{-1}= q_0^2 \,\cos 2q_0 \approx 10.1887$,
which satisfies the vacuum stability condition \eqref{eq:Compressibility}.

\subsection{Possible spacetime origin of $\boldsymbol{q}$: Aether velocity field}
\label{sec:Possible-spacetime-origin-aether}

Another possible origin of $q$ may be through
a four-vector field $u^{\mu}(x)$. This vector field could be the
four-dimensional analog of the concept of shift in the deformation
theory of crystals. (Deformation theory can be described in terms of a
metric field, with the role of  torsion and curvature fields played by
dislocations and disclinations, respectively;
see, e.g.,  Ref.~\cite{Dzyaloshinskii1980} for a review.)
However, a better realization of $u^{\mu}$ would be as a 4--velocity field
entering the description of the structure of spacetime.
In this case, one obtains a variant of the Einstein--aether theory discussed by
Jacobson \cite{Jacobson2007}, in which the timelike vector field is constrained
to have unit norm $u^{\mu}\,u_{\mu}=1$, or the more general vector--tensor
gravity theories studied by Will and Nordvedt \cite{WillNordvedt} (see also
Refs.~\cite{Gasperini1987,Jacobson2001}).

Here, we do not impose a constraint on the magnitude of $u^\mu$ and assume
that the action does not depend on $u^{\mu}$ explicitly but only
depends on its covariant derivatives $\nabla_\nu u^{\mu}\equiv u_\nu^\mu$.
This last assumption reflects the postulated Lorentz invariance of the
quantum vacuum: the vacuum does not depend on the
choice of the inertial reference
frame in which one has constant $u^{\mu}_{\nu}$ (see below).
Furthermore, we assume that the action contains higher order terms,
which allow for a nonzero value of $u_\nu^\mu$ in the equilibrium vacuum.
Specifically, the action is taken to have the following form:
\begin{equation}
S= S_\text{grav} +S_\text{vel}=
\int_{\mathbb{R}^4} \,d^4x\, \sqrt{-g}\,\left(\frac{R}{16\pi G}   -
\epsilon(u_\nu^\mu)\right)\,,
\label{eq:action3}
\end{equation}
with an energy density containing even powers of
$u_\nu^\mu\equiv \nabla_\nu\, u^{\mu}\,$:
\begin{equation}\label{eq:epsilon-u-mu-nu}
\epsilon(u_\nu^\mu)
= K
+ K_{\mu\nu}^{\alpha\beta}\,u_\alpha^\mu u_\beta^\nu
+ K_{\mu\nu\rho\sigma}^{\alpha\beta\gamma\delta}\,
u_\alpha^\mu u_\beta^\nu u_\gamma^\rho u_\delta^\sigma + \cdots \;,
\end{equation}
where the zeroth order term $K$
corresponds to a ``bare'' cosmological constant.
According to the imposed conditions, the
tensors $K_{\mu\nu}^{\alpha\beta}$ and
$K_{\mu\nu\rho\sigma}^{\alpha\beta\gamma\delta}$
depend only on $g_{\mu\nu}$ or $g^{\mu\nu}$ and the same holds for the other
$K$--like tensors in the ellipsis of \eqref{eq:epsilon-u-mu-nu}.
In particular, the tensor $K_{\mu\nu}^{\alpha\beta}$ of the quadratic term
in \eqref{eq:epsilon-u-mu-nu} has the following form in the notation of
Ref.~\cite{Jacobson2007}:
\begin{equation}
K_{\mu\nu}^{\alpha\beta}=c_1\, g^{\alpha\beta}g_{\mu\nu}  +
c_2\, \delta^{\alpha}_{\mu}  \delta^{\beta}_{\nu}  +
c_3\, \delta^{\alpha}_{\nu}  \delta^{\beta}_{\mu}  ~,
\label{eq:Kparameters}
\end{equation}
for real constants $c_n$ with mass dimension 2.
Distinct from the theory considered in Ref.~\cite{Jacobson2007},
our tensor \eqref{eq:Kparameters}
does not contain a term $c_4\, u^\alpha u^\beta g_{\mu\nu}$,
as such a term would depend explicitly on $u^{\mu}$ and
contradict our assumptions (motivated by the postulated Lorentz invariance
of the quantum vacuum).

The equation of motion for $u^{\mu}$,
\begin{equation}
\nabla_\nu  \, \frac{\partial\epsilon}{\partial u_\nu^\mu} =0\,,
\label{eq:Motion}
\end{equation}
has the solution expected for a vacuum-variable $q$--type field
in flat spacetime:
\begin{equation}
 \overline{u}_\nu^\mu=u\,\delta_\nu^\mu\,,\quad u=\text{constant}~,
\label{eq:solution3}
\end{equation}
where the specific solution of $u_\nu^\mu$ is indicated by a bar.
With this solution, the energy density in the action \eqref{eq:action3} is
simply $\epsilon(u)$
in terms of contracted coefficients $K$, $K_{\mu\nu}^{\mu\nu}$, and
$K_{\mu\nu\rho\sigma}^{\mu\nu\rho\sigma}$ from \eqref{eq:epsilon-u-mu-nu}.
However, just as in Sec.~\ref{sec:Possible-dynamic-origin-4form},
the energy-momentum tensor obtained by
variation over $g^{\mu\nu}$ and evaluated for solution \eqref{eq:solution3}
gives a \emph{different} energy density
denoted $\widetilde{\epsilon}_\text{vac}(u)$:
\beqa
\overline{T}_{\mu\nu}
&=&
\frac{2}{\sqrt{-g}}\;\frac{\delta S_\text{vel}}{\delta g^{\mu\nu}}\,
\Bigg|_{u_\nu^\mu=\overline{u}_\nu^\mu\,,\;g_{\mu\nu}=\overline{g}_{\mu\nu}}
 =               
\;\overline{g}_{\mu\nu}\left(\epsilon(u) - u\,\frac{d\epsilon(u)}{du}\right)
 \equiv
\overline{g}_{\mu\nu}\:\widetilde{\epsilon}_\text{vac}(u)\,,
\label{eq:emSolution3}
\eeqa
where $\overline{g}_{\mu\nu}$ stands for the flat spacetime metric.
This particular contribution to the energy-momentum tensor corresponds to the
following cosmological constant in Einstein's gravitational field
equations  \eqref{eq:classical-field-equations}:
\begin{equation}
\Lambda=       \epsilon(u) - u\,\frac{d\epsilon}{du}
       \equiv  \widetilde{\epsilon}_\text{vac}(u)\,,
\label{eq:CC3}
\end{equation}
whose origin was discussed in Sec.~\ref{sec:effective-vacuum-energy-density}
on general grounds.
In our formulation of the theory,
different from the Einstein--aether theory considered by Jacobson
\cite{Jacobson2007}, the nonzero vacuum value \eqref{eq:solution3}
of the field $u_\nu^\mu$ does \emph{not} violate Lorentz symmetry
but leads to compensation of the bare cosmological constant $K$
in the equilibrium vacuum.

However, the compensation
of the bare cosmological constant
occurs only for a vacuum solution with a nonzero equilibrium value $u_0$.
The quantum vacuum corresponding to the solution  $u_0=0$
would have a cosmological constant $\Lambda$ equal to the bare
cosmological constant $K\sim E_\text{UV}^4$.
This result illustrates an important difference
between the following
two kinds of vacua. The quantum vacuum with $u_0=0$, on the one hand,
can exist only with external pressure $P =-K$ and
is not a self-sustained medium (see also Footnote~\ref{ftn:selfsustained}).
By analogy with condensed-matter physics,
this kind of quantum vacuum may be called ``gas-like.''
The quantum vacuum with nonzero $u_0$, on the other hand,
can be stable at $P=0$, provided that a nonzero solution $u_0$ exists
for $P=0$ and that it has positive compressibility, $\chi(u_0)>0$.
This kind of quantum  vacuum may then be called ``liquid-like''
and a numerical example of such a vacuum
(with the nonzero equilibrium vacuum variable $q_0$ replacing the $u_0$ from here)
has been given in the last paragraph of the previous subsection.

\subsection{Possible spacetime origin of $\boldsymbol{q}$: Minimal volume}
\label{sec:Possible-spacetime-origin-minimal-volume}

In this subsection, another suggestion for the possible origin
of the vacuum variable $q$ is made
(mathematically, this suggestion is similar to the one
of Sec.~\ref{sec:Possible-dynamic-origin-4form} but, physically, not).
The relevant variable $q$ might,
in fact, come from a fundamental anti-symmetric
field such as $q_{\mu\nu\rho\sigma}=q \,e_{\mu\nu\rho\sigma}$.
This $q$ would then be related to some kind of universal
minimal 4--volume $l^4$ (cf. Sec.~4.11 of Ref.~\cite{Weinberg1972}),
so that $q= \widehat{q}\:l^4$ for a dimensionless number $\widehat{q}$.
This minimal volume $l^4$ may be essential
to the underlying microscopic theory \cite{Klinkhamer2007},
but it is the dimensionless vacuum variable $\widehat{q}$ which adjusts
itself to zero pressure, $P=0$, and this adjustment
leads to a vanishing effective vacuum
energy density, $\widetilde{\epsilon}_\text{vac} = - P = 0$.

Let us end this section with a general remark.
The separation of Secs.~\ref{sec:Possible-dynamic-origin-general} and
\ref{sec:Possible-dynamic-origin-4form} in one group and
Secs.~\ref{sec:Possible-spacetime-origin-aether} and
\ref{sec:Possible-spacetime-origin-minimal-volume} in another group
has been made in order to simplify the discussion.
However, it may very well be that the origin of $q$  lies
in a fundamental theory
which combines ``dynamic fields'' and ``spacetime structure,''
as alluded to in the first paragraph of the Introduction.

\section{Summary and outlook}
\label{sec:Summary}

In this article, we have suggested that the value of
the cosmological constant $\Lambda$
is determined by a self-tuning thermodynamic variable $q$
of the Lorentz-invariant self-sustained quantum vacuum.
The natural value of the gravitating vacuum energy density
$\widetilde{\epsilon}_\text{vac}(q_0)$, identified with the cosmological
constant $\Lambda$, would then be zero.
For the perfect quantum vacuum, the equilibrium value $q_0$
adjusts itself so that $\widetilde{\epsilon}_\text{vac}(q_0)=0$.
In our considerations, a crucial role is played by the Lorentz invariance
of the quantum vacuum, for which there is strong experimental support.

A small nonzero value of the cosmological constant
(that is, small compared to Planck-scale energy densities)
may result from perturbations
of the Lorentz-invariant state of the universe,
for example, by the presence of thermal matter. In fact,
the presence of matter violates Lorentz invariance by introducing
a preferred reference frame, where the matter is, on average, at rest.
The resulting state of the universe is no longer Lorentz invariant
and this allows for having a nonzero value of the vacuum energy density
$\widetilde{\epsilon}_\text{vac}$.
However, the value of the vacuum energy density is small,
as it is proportional to the perturbation which violates
the original Lorentz invariance of the perfect (unperturbed) quantum vacuum.

The implication of our suggestion is that
the so-called ``dark energy'' would be homogeneous,
provided the matter is distributed homogenously over large scales.
But the precise interaction of gravitating matter
(visible and ``dark'') with the microscopic degree of freedom
$q$ from the quantum vacuum remains to be clarified.
Perhaps, carefully planned astronomical observations can give a clue.

The vacuum energy density is governed
by processes in the deep ultraviolet vacuum.
Still, the approach followed in this article demonstrates
that the thermodynamics of the vacuum energy can be described by
a relatively minor extension of the low-energy (infrared)
effective theory, namely, by the introduction of a \emph{constant}
vacuum variable $q$. This particular model expands on the idea
that the cosmological constant problem belongs to the
realm of infrared physics \cite{Veltman1976,Polyakov1982}.

The introduction of an infrared thermodynamic variable $q$,
describing certain properties of the deep vacuum
and having the equilibrium value $q_0 \equiv q_\text{vac}$,
can be considered as the first step in a bottom-up approach
(trying to go from the effective low-energy theory to the
fundamental microscopic theory). This simple approach already allowed us
to introduce the thermodynamic notion of the vacuum compressibility
$\chi_\text{vac}$ which is a new fundamental constant for the
low-energy world and to estimate the thermodynamic back reaction
on these constants ($q_\text{vac}$, $\chi_\text{vac}$, and others)
by low-energy phase transitions
and the presence of thermal matter.

In this article, we have studied what may be called
the ``statics of dark energy.''
It is to be hoped that the \emph{dynamics} of the vacuum energy density,
which describes the relaxation of the ``cosmological constant''
to its equilibrium value,
can be obtained  by a further extension of our effective theory.

\section*{\hspace*{-4.5mm}ACKNOWLEDGMENTS}

It is a pleasure to thank M.J.G. Veltman and A.M. Polyakov for helpful
discussions. GEV is supported in part by the Russian Foundation for Basic
Research (grant 06--02--16002--a).

\end{document}